\documentclass{article}

\begin{document}

\title{Relative Divergence and Maximum Relative Divergence Principle for Grading Functions on Partially Ordered Sets}

\author{Alexander Dukhovny\\
\small{Department of Mathematics, San Francisco State University} \\
\small{San Francisco, CA 94132, USA} \\
\small{dukhovny@sfsu.edu} \\ }

\date{\today}          

\maketitle

\begin{abstract}
    
    Relative Divergence (RD) and Maximum Relative Divergence Principle (MRDP) for grading (order-comonotonic) functions (GF) on posets are used as an expression of Insufficient Reason Principle under the given prior information (IRP+). Classic Probability Theory formulas are presented as IRP+ solutions of MRDP problems on conjoined posets. RD definition principles are analyzed in relation to the poset structure. MRDP techniques are presented for standard posets: power sets, direct products of chains, etc. "Population group-testing" and "Single server of multiple queues" applications are stated and analyzed as "IRP+ by MRDP" problems on conjoined base posets.
    
\end{abstract}

\section{Introduction}
\label{sec:Intro}

In many probability theory problems Maximum Shannon Entropy Principle (MEP) has been used as an expression of Insufficient Reason Principle (IRP) to choose the "least-presuming" IRP solutions by maximizing, under the application's constraints, Shannon Entropy (SE) functional (see, e.g., \cite{Shannon}, \cite{Jaynes}).

The original SE formula and MEP have been restated in many ways: Kullback-Leibler (K-L) Relative Entropy and Divergence, Partition Entropy, General (non-additive, fuzzy, etc.) Measures, and many others - see, e.g., the review in \cite{Review}, our paper \cite{Dukh} and \cite{Marichal}, \cite{KMR}, \cite{KM}, \cite{HG}).

The approach of \cite{Dukh} and \cite{AOME} is continued here to extend SE and K-L entropy concepts to that of Relative Divergence (RD) of order-comonotonic grading functions (GF) on partially ordered sets (posets). (The term Relative Divergence itself comes from Kullback-Leibler Divergence of probability distributions, see (\cite{K-L}).)

Based on that, we generalize MEP to MRDP (Maximum Relative Divergence Principle) as a method to obtain the least-presuming "IRP+" solutions for cases where models are stated in terms of GFs on posets and some prior information ("null" GF, inherent constraints, GF admissibility requirements, etc.) is present. In such problems MRDP is used here as a mathematical expression of IRP+. As such, it can be used as a tool for  updating the "null" model based on the prior information to the new one reflecting the new data in the least-presuming way.)

In section \ref{Basics} we begin with definitions and basic properties of RD and MRDP on totally ordered chains as introduced in our \cite{AOME}. In particular, using MRDP as an expression of (IRP+) leads to the classic conditional probability formula.

In section \ref{Posets} we move on to cases where grading functions are defined on a partially ordered set (poset) $W$. We discuss fundamental issues in the setup of the application's model as an MRDP problem on its conjoined base poset: the poset structure, the ordering relation, the nature of the grading functions and possible constraints.

In particular, the standard formula for probabilities of independent events is derived using MRDP as n expression of IRP+ on the conjoined base poset constructed for that purpose.

For some special poset structures emerging in applications (say, circuits in neural/electrical networks) we introduce "additivity" and "infinum" principles for assembling the RD on the entire circuit from RDs computed on its components. 

Based on that, in section [\ref{MRDP on Psets}] MRDP is applied to the power set $W = 2^X$ of an "element space" $X$, $W$ ordered by subset inclusion. Kullback-Leibler Divergence is generalized to probability distributions on power sets, and Partition Entropy concept is presented as a reduction of RD to the power set of the space of the partition subsets.

As an example, we explore a "Population group-testing" application where the tested groups form a partition of the population $X$. MRDP analysis is applied as the IRP+ tool both when no prior information is present on  the group-testing cost function and when new actual group-cost data appears to update that function in the least-presuming way to facilitate further planning decisions.

In section [\ref{RD and MRDP on Bundles}] $W$ is a direct product ("bundle") of chains, under the standard direct product order, as presented and studied in \cite{MRDP on Bundles}. Direct results are obtained in cases where grading functions can be assumed to have special properties: "height-dependent" or "additively separable". 

For example, we explore a Queueing Theory application: a service batch is formed from separate queues. Using MRDP, the conjoined base poset $W$ is a bundle of those queues. MRDP is applied to analyze the batch service cost and the least-presuming probability distribution of the "server's type" when the new information suggests that the service "cost" is type-dependent.

\section{Basic Definitions and Properties}\label{Basics}

The initial RD setup (see \cite{AOME}) begins as follows: let $W = \{ w_k, \quad k \in Z,$ be a "chain" - a set totally ordered by a relation $\prec$. A real-valued function $F$ on $W$ is said to be a grading function (GF) on $W$ if it is order-comonotonic, that is,

$w \prec v \iff F(w) < F(v)$ for all $w, v \in W $.

(Say, the "indexing" function $I: I(w_k) = k$ is a "natural" GF on $W$) .

For GFs $F(w)$ and $G(w)$ defined on $W$, the RD of $F$ from $G$ on $W$ is defined (assuming absolute convergence of the series), as

\begin{equation} \label{RDcountable}
\mathcal{D}(F \Vert G) \vert_W = -\ \sum_{k \in Z}
\ln  \left( \frac{f_k}{g_k}\right) f_k, 
\end{equation}

\noindent where
$f_k=\Delta_k F = F(w_k) - F(w_{k-1})$, \quad $g_k=\Delta_k G = G(w_k) - G(w_{k-1}),$

\noindent are increments of both GFs along the chain $W$.

In the special case where $F$ is a CDF (probability cumulative distribution function) on the chain $W$ and $G = I,$ definition (\ref{RDcountable}) reduces to:

\begin{equation} \label{RD-Shannon}
\mathcal{D}(F \Vert I) \vert_W = -\ \sum_{k \in Z} f_k \ln {f_k}, 
\end{equation}

\noindent -showing the connection of RD of GFs on $W$ and Shannon Entropy $\mathcal{H}(F)$ of $F$.\\

Some linearity properties of $\mathcal{D}(F \Vert G) \vert_W$ follow directly from its definition.

1. $\mathcal{D}((c+F) \Vert (d+G)) \vert_W = \mathcal{D}(F \Vert G) \vert_W . \quad \forall c, d$

2. $\mathcal{D}(cF \Vert cG) \vert_W =
c\mathcal{D}(F \Vert G) \vert_W, \quad \forall c>0.$

3. Where $g(e) = 1, \forall e \in E(W)$ and the value range of $F$ is $[ m, M ]$

$\mathcal{D}(cF \Vert G) \vert_W =
c\mathcal{D}(F \Vert G) \vert_W - (M-m)c\ln{c}, \quad \forall c>0.$\\

In particular, defining the "normalized" $F(w)$ as $\hat{F}(w) = \frac{F(w) - m}{M-m},$ it now follows that General Entropy of $F$ on $W$ differs from Shannon Entropy of $\hat{F}$ by a constant
 
$\mathcal{D}(F \Vert N) \vert_W = 
\mathcal{D}(\hat{F} \Vert N) \vert_W  + \ln{(M-m)}$,

\subsection{Maximum Relative Divergence Principle for grading functions on chains}

The Maximum Relative Divergence Principle (MRDP) for GFs on chains - a generalization of the Maximum Entropy Principle (MEP) - is introduced as follows:

MRDP: Among all admissible GFs on a chain $W$ with the same values interval, $F$ is said to be the "least-presuming" if its RD from the given "null" GF $G$ is the highest possible.

\newpage

The connection between Shannon Entropy and RD of grading functions makes some classic results from the standard toolkit of Shannon Entropy theory (see, e.g., \cite{Jaynes}) relevant here:.

\textbf{Lemma One} For a probability distribution 

$\{p_i\}, \quad 0\leq p_i\leq 1, \quad \sum_{i=1}^n p_i = 1, \quad i = 1, \ldots, n$,

\noindent the maximum value of Shannon's Entropy of that distribution

$\mathcal{H}  = -\sum_{i=1}^n p_i\ln{p_i} = \ln{n}$

\noindent is attained when $p_i = \frac{1}{n}, \quad i = 1, \ldots, n$.\\ 

A useful result follows directly from \textbf{Lemma One} (see also \cite{MRDP on P-sets}):

\textbf{Proposition 2.1.1} Let $F(i)$ be a GF on the chain $W = \{0, \ldots, n \}$, and suppose some values of $F(i)$ are fixed: $F(n_k)=m_k, \quad k=1, \ldots, K$, 

\noindent where $n_0 = 0, \quad m_0 = m,\quad n_K = n, \quad m_n = M$ 

\noindent and \quad 
$\Delta_k m = m_k - m_{k-1}$, \quad $\Delta_k n = n_k - n_{k-1}$, \quad $k = 1, \ldots, K.$

\noindent Then the maximum value of 
$\mathcal{D}(F \Vert I) \vert_W = \sum_{k=1}^K [\Delta_k m \Delta_k n)- \Delta_k m \ln \Delta_k n)]$

\noindent is attained when $F$(i) is a piece-wise linear function:

\begin{equation} \label{Proposition 2.1.1 formula}
F(i)=a_k+b_k i, \quad i \in I_k, \quad k=1, \ldots, K, \quad \forall i \in W,
\end{equation}

\noindent where 
$\quad b_k= \frac{ \Delta_k m} {\Delta_k n}, \quad a_k = m_k - b_k n_{k-1},$,  $I_k = ( n_{k-1}, n_k ], \quad k = 1, \ldots, K$.\\

In general, to use MRDP analysis of an application whose model is stated on a base chain, it must be stated in terms of GFs defined on that chain, and its natural structural conditions and constraints need to be formulated accordingly. Based on the pre-update information, the "natural" null GF needs to be stated as well.

\subsection{Conditional Probability formula by IRP+ expressed as MRDP}

As an example of using MRDP as an expression of IRP+ the Conditional Probability formula Theory is derived here (see \cite{P(A|B)}):

$P(B|A) = \frac{P(A \cap B)}{P(A)}$.

\textbf{Proof}. In a random experiment with the outcome space $U$, set event inclusion as an ordering relation $\prec$ on $U$. Use a set of events $W = \{ \emptyset , A\cap B, A \}$ as the base chain. Let function $G$ on $W$ have values, respectively, $\{ 0, p_1 = P(A \cap B), p_2 = P(A)$, and function $F$ have values $0, x = P(B|A), 1$. As such, $G$ and $F$ are GFs on $W$. 

When only $p_1$ and $p_2$ are known, by IRP+ $G$ must be set as the "null" GF on $W$, whereas $F$ is to be specified by the value of $x$. Using MRDP for IRP+ calls for finding $x \in (0,1)$ to maximize  $q(x) = \mathcal{D}(F \Vert G) \vert_W.$ By (\ref{RDcountable}), 

$q(x) =- x\ln \frac{x}{p_1} - (1-x)\ln \frac{1-x}{p_2 - p_1},$

$q'(x) = \ln{(1-x)} - \ln{x} + \ln{p_1} - \ln{(p_2 - p_1)} $

$q''(x) = -{[(1-x)x]^{-1}} < 0,$

\noindent so $q(x)$ is concave down on $(0, 1)$ and has its only maximum there at $x = \frac {p_1}{p_2}$, proving the Conditional Probability formula.

\newpage

\section{MRDP on general posets}\label{Posets}

Applying MRDP to problems involving GFs whose domain $W$ is a poset, the setup of the RD and MRDP has to take into account the structure of the order relation on $W$, the nature of the GFs suggested by the application, and admissibility constraints. Therefore, our approach here is to apply our previous results to each maximal chain $C \in W$, where the GFs on $C$ are reductions to $C$ of GFs defined on the entire $W$. Accordingly,  definitions of RD and MRDP have to take into account applications-specific interdependence of those maximal chains and MRDP problems on them.

\subsection{MRDP for probabilities of independent events}\label{P(AB)}

Example: the MRDP setup for deriving the Independent Events Probability formula from IRP+ (see \cite{P(AB)}

Consider a random experiment with the outcome space $U$ where events $A, B, A\cap B, A\cup B$ are defined. In the framework of IRP+, to decide on independence of $A$ and $B$ their probabilities $p_1 = P(A)$ and $p_2 = P(B)$ must be specified - but no other information is to be presumed.

\textbf{Proposition 3.1.1}. When the only information on the relation of events $A$ and $B$ consists of $p_1 = P(A)$ and $p_2 = P(B)$, using MRDP as an expression of IRP+ leads to the classic formula: 

\begin{equation}\label{independence}
P(A\cap B) = P(A)P(B)  
\end{equation}

\textbf{Proof}. Let $W = \{ \emptyset, A\cap B, A, B, A \cup B, U \}$ with event inclusion as the ordering relation $\prec$.  As such, $W$ is a poset. The only maximal chains of events in $W$ are
$C_1 = \{ \emptyset, A\cap B, A, A\cup B, U \}, \quad C_2 = \{ \emptyset, A\cap B, B, A\cup B, U.\}$\\

With no information on the probabilities of the events of $U$, by IRP+ the only "null" GF defined on both chains is the index function $G$ whose values on both $C_1$ and $C_2$ are $\{ 0, 1, 2, 3, 4 \}$, and its  increments along both $C_1$ and $C_2$ are $g_i = 1, i= 1,2,3,4$. 

When the new information provides only $p_1$ and $p_2$, GFs $F_1$ and $F_2$ on the chains $C_1, C_2$ are assigned their values based on that information and probability axioms:

$F_1(\emptyset) = F_2(\emptyset) = 0, \quad F_1(U) = F_2(U) = 1$,

$F_!(A) = p_1, \quad F_2(B) = p_2$

\noindent Denoting the unknown $P(A\cap B) = x$, by the probability axioms, we assign 

$F_1(A\cap B) = F_2(A\cap B) = x,$ \quad and

$F_1(A\cup B) = F_2(A\cup B) = P(A\cup B) = p_1 + p_2 - x$,

\noindent completing the definitions of $F_1$ and $F_2$,

By its definition, $x \in {[max(0, p_1+p_2 - 1), min(p_1, p_2)]}$. 

Now we follow (\ref{RDcountable}) to compute the RDs 

$d_1(x) = \mathcal{D}(F_1 \Vert G) \vert_{C_1} $  and  $d_2(x) = \mathcal{D}(F_2 \Vert G) \vert_{C_2} = d(x) =$ 

\noindent $ -x\ln{x}-(p_1-x)\ln{(p_1-x)} - (p_2-x)\ln{(p_2-x)} - (1-p_1-p_2 + x)\ln{(1-p_1-p_2 + x)}.$

Using MRDP as an expression of IRP+, we find the $x$ that maximizes $d(x)$ on its domain - the interval $(max[0, p_1+p_2 - 1], \quad min[p_1, p_2])$.  Now,

$d'(x) = - \ln{x} +\ln{(p_1-x)} + \ln{(p_2 - x)} - \ln{(1-p_1 - p_2 +x)}, $

\noindent and its only root in its domain is $x = p_1 p_2.$  Next,

$d''(x) = -\frac{1}{x} - \frac{1}{(p_1 - x)} - \frac{1}{(p_2 - x)}- \frac{1}{(1-p_1 - p_2 +x)} < 0,$

\noindent that is, $d(x)$ is concave down on its domain, so the only maximum value of $d(x)$ there is attained at 
$x = p_1p_2$,  concluding the proof of Proposition 3.1.1.\\

Mathematical models of some applications may have application-specific features that can be used to simplify their MRDP analysis.

Say, where the base poset $W = \cup W_k, \quad k=1,\ldots, K,$ is a union of disjoint components and the application-admissible GFs are defined separately and independently on each $W_k$, the overall MRDP analysis simply breaks down into separate studies of individual components.

When the components are "order-disjoint" (that is, their edge-sets $E(W_k)$ are disjoint but may have common points) values of admissible GFs at common points become component-linking constraints and complicate the overall analysis.

For example, suppose $W = \cup W_k, k=1,\ldots, K$ where all components have a single common point  Then MRDP can still be applied individually to each component - but the optimal GFs for each component must have their values additively shifted by suitable constants in order to agree at the common point.

Generally, the diversity of types of partial order relations on the base poset emerging in application models may not allow for one universal definition of Relative Divergence and MRDP on posets. In this paper we explore cases arising in some important applications.

\subsection{"Lowest-to-greatest" posets}\label{l-g posets}

A poset $W$ is said to be an [$l-g$] poset if it contains its "lowest" and "greatest" elements $l(W)$ and $g(W):$ \quad  $l(W) \preceq s \preceq g(W), \quad \forall s \in W$.\\

\textbf{Proposition 3.3.1} A [$l-g$] poset us a union of [$l-g$] maximal chains.

\textbf{Proof}. By definition, $\forall s \in W$ there have to be chains of ordering edges both from $l(W)$ to $s$ and from $s$ to $g(W)$. Those chain form a maximal chain in $W$.\\

\textbf{Proposition 3.3.2} Any poset $W$ can be embedded into a unique minimal [$l-g$] poset $\hat{W}$ - a minimal [$l-g$] order-enclosure of $W$.

\textbf{Proof}. Augment $W$ with added $l$ and $g$ elements (if not already present in $W$), and add to $E(W)$ $\prec$ ordering edges running, respectively, from $l$ to all minimal, and to $g$ from all maximal elements of $W$. Now, each $w \in \hat{W}$ is either $l$ or $g$ , or is connected to both by the edges of $\hat{W}$, so $\hat{W}$ is, indeed, an [$l-g$] poset.\\ 

A GF $F(w)$ can be extended from $W$ to $\hat{W}$ (preserving its range of values) by setting 
$F(l) := \inf_{w \in W} F(w), F(g) := \sup_{w\in W} F(w).$\\

In this paper we apply MRDP analysis to posets made of "blocks": ordered [$l-g$] posets.

A poset $W$ is a "block-chain" if it is a union of concatenated blocks $W_1 \cup \ldots \cup W_K$ where  $g(W_k) = l(W_{k+1}), \quad k = 1, \ldots, K-1$ ("serial connection").

A poset $W$ is a "block-split" if it is a union of blocks $W = W_1 \cup \ldots \cup W_K$ with the same endpoints $l,g$ ("parallel connection").

In both cases the ordering relation $\prec$ on $W$ is imposed by the $\prec$ directed edges of $E(W) = E(W_!) \cup \ldots \cup E(W_K)$, so $W$ is an [$l-g$] poset.\\

In particular, a "split-chain" $C$ is a union of [$l-g$] chains: $C = \cup C_l, l = 1,\ldots, L$ all with the same endpoints. A split-chain $C$ is said to be "even-sided" when all its component chains have the same number of edges: \quad $|E(C_1)| = \ldots = |E(C_L)|$,\\

Definitions of RD on block-chains and block-splits are based on the following general principles.\\

\textbf{RD block-additivity on block-chains:}

\begin{equation} \label{block-chain RD}
\mathcal{D}(F \Vert G) \vert_W = \sum_{k=1}^K \mathcal{D(F \Vert G)} \vert_{W_k},
\end{equation}

\textbf{RD chain-infinum on even-sided split-chains:}

\begin{equation} \label{e-s split-chain RD}
\mathcal{D}(F \Vert G) \vert_C = \inf_{\forall l} \mathcal{D(F \Vert G)} \vert_{C_l},
\end{equation}
.
When the application-suggested definition of Relative Divergence of GFs on the base poset $W$ is constructed, we follow \cite{AOME} in expressing Insufficient Reason Principle IRP+ by MRDP stated as follows.\\

\textbf{MRDP}: Among all application-admissible GFs on $W$ with the same range of grades, $F$ is said to be IRP+ suggested ("least-presuming") if its RD from the given "null" GF $G$ is the highest.\\

In the special case where the "null" edge-function $g(e) = 1, \forall e \in E(W)$, $G$ is said to be a "natural" GF $N$ on $W$. In that case we retain the term "General Entropy" for 
$\mathcal{H}(F) \vert_W = \mathcal{D}(F \Vert N) \vert_W.$

\subsection{MRDP analysis on power sets}\label{MRDP on Psets}

In \cite{MRDP on P-sets} MRDP analysis was applied to the case where $W = 2^X$, the powerset of a set $X$, ordered by subset inclusion. It continued the work started in \cite{Dukh}, where the grading functions were "general" (non-additive) normalized measures, so the Relative Divergence formula was linked there to that of K-L Relative entropy of probability distributions (see \cite{K-L}).

Specifically, the values of a general measure $\mu$ on any maximal chain $C \in W$ were used to (uniquely) specify an additive "subordinate" measure $\mu _C$. The General Entropy of $\mu$ on $W$ was then defined there as the infinum of Shannon Entropies of all measures $\mu_C$ over all maximal chains in $W$. In \cite{MRDP on P-sets} that method was further extended  to define RD of GFs $F$ and $G$ on $W$ as a foundation for MRDP.

Here we show that those results follow directly from (\ref{e-s split-chain RD}).

Indeed, each maximal chain $C \in W=2^X$ consists of nested subsets of $X$ differing by a single extra-element. As such, it is uniquely specified by a permutation of the elements of $X$, begins with $l=\emptyset$ and ends with $g=X$ and has the same number of edges $L=|X|$. Therefore, $W$ is an even-sided split chain, and for any GFs $F$ and $G$ on $W$ by (\ref{e-s split-chain RD})

\begin{equation} \label{RD on 2^X}
\mathcal{D}(F \Vert G) \vert_C = 
\inf_{\forall MC \in W} \mathcal{D(F \Vert G)} \vert_{MC},
\end{equation}

In particular, with no prior information available, the subset-cardinality function $N(w) = |w|, \forall w \subseteq X$, is the natural null function $G$, the term General Entropy of $F$ is retained.

\subsection{MRDP for subset-cardinality dependent grading functions}
\label{MRDP for C-D case}

In numerous applications an admissible grading functions $F(w)$ on $W = 2^X$ must be "subset-cardinality dependent", that is,

$F(w) = F(|w|, \quad \forall w \in W, \quad F(0) = 0, \quad F(|W|) = M$.

Therefore, applying MRDP to admissible grading functions obviously requires only one maximal chain.

When the only constraints on $F(w)$ are that $F(\emptyset) = 0, F(X) = M$ and $N(w) = |w|$ is used as a "null" grading function, it follows directly from \textbf{Lemma One} that the least-presuming (by MRDP) $F(w)$ is

$F(w) = M\frac{|w|}{N}$.

If, in addition, new constraints require that some values of $F(|w|)$ should be prespecified, that is, $F(n_k) = M_k, \quad k=1, \ldots, K$, 

\noindent defining $n_0 = 0, \quad M_0 = m,\quad n_K = n = |X|, \quad M_K = M$,

\noindent and index intervals $I_k = ( n_{k-1}, n_k ], \quad k = 1, \ldots, K$

It follows directly from \textbf{Proposition 2.1,1} that the overall solution of the MRDP problem presents as a piece-wise linear function as follows:

\begin{equation} \label{C-D solution)}
F(w)=a_k+b_k |w|, \quad |w| \in I_k, \quad k=1, \ldots, K, \quad \forall w \in W
\end{equation}

\noindent where
$\quad b_k= \frac{ M_k - M_{k-1} }{n_k - n_{k-1}}, \quad a_k = M_k - b_k n_{k-1}.$

\subsection{MRDP for "partition-induced" Grading Functions}
\label{MRDP for p-i case}

Another important application of RD and MRDP on a power set is related to what is known as "Partition Entropy" (see, e.g., \cite{Sinai}).

Consider a set $S$ whose elements are  disjoint subsets of $X$ comprising a partition of $X$: $\mathcal{S} = \{ s_1, s_2, \ldots \}$.   

Let $f(s)$ and $g(s)$ be positive-valued functions defined on $S$ and let

$F(v) = \sum_{\forall {s_k \in v}} f(s_k), \quad  
G(v) = \sum_{\forall {s_k \in v}} g(s_k),$

\noindent for any subset $v \subseteq S$.

By construction,  both $F$ and $G$ are grading functions on $S$, so equation (\ref{RD on 2^X}) gives rise to the concept of "partition-induced" Relative Divergence:

\begin{equation}\label{eq:(partition-induced RD)}
\mathcal{D}(F \Vert G)\vert_S  = - \ \sum_{\forall k} f(s_k)  \ln \left( \frac{f(s_k)}{g(s_k)} \right)
\end{equation}.

When $g(s_k) = 1, \forall k$, that equation reduces to

\begin{equation}\label{eq:(partition entropy)}
\mathcal{D}(F \Vert G)  = - \ \sum_{\forall k} f(s_k)\ln({{f(s_k)}}), 
\end{equation}

\noindent which is known as Partition Entropy for the special case where $f$ comes from a probability distribution $p(x)$ on $X$: $f(s_k) = \mathcal{P} (x \in s_k) =  \sum_{\forall x \in s_k} p(x) $.

When the partition subsets are just single elements of $X$, and both functions $f(x)$ and $g(x)$ represent probability distributions on the outcome space $X$, formula (\ref{eq:(partition-induced RD)}) extends the classical formula of Relative Entropy (see, e.g., \cite{K-L} of those probability distributions to the case of a general outcome space $X$.

\textbf{Representative example}: testing subjects from a population $X, |X| = N$ for all "positive" subjects. Tests can be done either individually, one-by-one, or by first applying a special preliminary collective pretest to the entire sampled group. If the pretest is negative, all subjects in the group are certified negative. Otherwise, the group is partitioned further into smaller groups, which are tested separately, and so on. (See, e.g., HIV-testing methods explored in (\cite{AbDu})

Typically, the cost $G$ of testing a group $w$ increases with the groups size $|w|$, making $G(w)$ a GF on $W=2^X$. At the earliest planning stage the available information may consist of just the statistical estimate of the population's "attribute" incidence, size $N$ and the maximum budgeted "cost" $M$.  As such, the null $G(w)$ may be assumed depending only on the group size $|w|$. Now, by \textbf{Lemma One}, the "least-presuming" null $G(w) = |w|\frac{M}{N}$.

Should more data on testing cost emerge, the null $G(|w|)$ must be updated accordingly. Say, if some of its values have to be fixed, then, using MRDP, the "least-presuming" formula of Proposition 2.1.1 can be  used as the new null $G(w)$.

Now, testing the entire population $X$, the subpopulations test groups $s_k, k=1, \ldots,K$ form a partition of $X$.The newly obtained $G(w)$ can be used to update the testing cost of each $s_k$ as $G(s_k)$, thereby creating a basis for further planning decisions.

Conversely, suppose the new information, statistical or mandated, does not support the "subset- cardinality-dependence" as the "null" assumption. Instead, it  specifies group-testing cost of some observed groups. Updating the null $G(w)$ in the least-presuming way to the new $F(w)$, MRDP can now be applied under constraints imposed by the specified values.

\section{RD and MRDP on Chain Bundles}
\label{RD and MRDP on Bundles}

In \cite{MRDP on Bundles} we studied a special case where $W$ is a "chain bundle" - a direct product of $K$ totally ordered chains: 
 
$W = X_1 \otimes \ldots \otimes X_K$, where  $X_k = \{ x_k (i), \quad i = 0, 1, \ldots, n_k \}, \quad k = 1, \ldots, K$.

The standard order relation on $W$ is imposed by the order of the elements' vector indices: for all unequal vectors in $W:$
 
$\vec{w}_{\vec{i}} \prec  \vec{w}_{\vec{j}} \iff \vec{i} \prec \vec{j}$, \quad (that is, $i_k \leq j_k, \quad \forall k = 1, \ldots, K$).

By construction, $W$ is an [$l-g$] poset: 

$l = \vec{w}_0 = [x_1 (0), \ldots, x_K (0)]$, \quad $g = \vec{w}_Q = [x_1 (n_1), \ldots, x_K (n_K)]$.

In vector notation, $W = \{ \vec{w} (\vec{i}) \} $, its elements are $ \vec{w} (\vec{i}) = [x_1 (\vec{i}), \ldots, x_K (\vec(i)]$, and their vector indices $\vec{i} = [i_1, \ldots, i_K]$.
 
Since the definitions of elements of $W$ do not involve their actual values, where feasible, the element $\vec{w} (\vec{i)}$ will be referred to simply by its index vector $\vec{i}$. Also, the element's "height" is defined as

$N(\vec{w} (\vec{i))} = N(\vec{i}) = i_1+ \ldots +i_K $

Each maximal chain $MC$ in $W$ consists of adjacent elements whose vector indices differ by 1 in only one of its components. Therefore, all maximal chains in $W$ have $Q = n_1 + \ldots + n_K$ elements, so $W$ is an "even-sided split-chain"

For a GF $F(\vec{i})$ defined on $W$ we also define 

$f_{MC} (\vec{i}(q)) = 
F(\vec{i} (q)) - F(\vec{i} (q-1)), \quad q = 1, \ldots, Q,$ 

\noindent - the "increment" function of $F$ along the chain $MC$. As a GF, $F$ is order-comonotonic, so $f$ is nonnegative. We also denote $m = F(l), M = F(g)$.

If the application does not impose a null GF on the chain bundle $W$, IRP-suggests element's height $N(\vec{i})$  function whose increment function $f_{MC} (\vec{i}(q)) = 1, \quad q = 1, \ldots, Q$ along any maximal chain $MC$.

In \cite{MRDP on Bundles} we applied MRDP analysis to applications where mathematical models impose additional constraints on the admissible null or updated GFs, both on their values and structure.
\\

\textbf{"Height-dependent" Grading Functions on Chain Bundles}

A GF $F$ on a chain bundle $W$ is said to be  "height-dependent" if $F(\vec{i}) = F(N(\vec{i}))$. In that case, when the minimum and maximum $M$ costs $m, M$ are fixed. formula (\ref{MRDP for F(N)}) yields the "least-presuming" GF (see \cite{MRDP on Bundles}):

\begin{equation} \label{MRDP for F(N)}
F(\vec{i}) = m + N(\vec{i}) \frac{(M-m)}{Q}.
\end{equation}

and, via Proposition 2.1.1 and formula (\ref{MRDP for F(N)}), the maximum possible value of

$\mathcal{D}(F \Vert N) \vert_{W} = (M-m)\ln Q - (M-m)\ln{(M-m)}.$
\\

\textbf{"Additively Separable" Grading Functions on Chain Bundles}

A GF $F$ is said to be "additively separable" on $W$ if $F(\vec{i}) = \sum_{k=1}^K F_k(i_k),$. (Say,
the "height" function $N(\vec{i})$ is additively separable by its definition.)

In \cite{MRDP on Bundles} the following proposition has been presented and proved .

\noindent \textbf{Proposition 4.1.1} If both GFs $F$ and $G$ on $W = W_1 \otimes \ldots \otimes W_K$ are additively separable, then

\begin{equation} \label{RD all +separable}
\mathcal{D}(F \Vert G) \vert_W = 
\sum_{k=1}^K \mathcal{D}(F_k \Vert N_k) \vert_{W_k}.
\end{equation}

A representative case arises in the Queuing Theory context: a service system is observed where a server forms a service batch by combining groups of customers selected from $K$ separate totally ordered queues. Each such group $i_k$ is, therefore, an element of a chain $W_k$ of nested subsets of the $k$-th queue. As such, the service batch $\vec{i}$ can be modeled as an element of the bundle $W=W_1 \otimes\ldots \otimes W_K$ of those chains - a "bundle of queues".

The mathematical model here has to reflect the service contract in the following aspects.

1. Specifying the set of acceptable  batches. (Say, depending on the server's overall and/or queue-specific capacities.) With no prior information constraining the service batch composition the natural "null" assumption is that the entire $W$ is the base poset of the model.

2. Specifying the calculation formula of the total batch service "cost" $F(\vec{i})$ in relation to the batch composition and service priorities. With no prior information to the contrary, IRP suggests that group selections from the component queues may be assumed independent. 

3. Specifying the way the total batch service cost depends on its composition.\\

In the simplest case, where it is plausible to assume that cost determined by the batch size alone, the "height-dependent" solution formula (\ref{MRDP for F(N)}) can be used as a "least-presuming" solution.

Otherwise, where the service costs differ among individual queues, the total batch service "cost" $F$ is often modeled as the sum of the "batch forming" cost and individual service costs of all selected groups. Then the formula of  (\ref{RD all +separable}) can be used to break down the overall MRDP analysis into individual problems for "least-presuming solutions" for each queue.

\subsection{MRDP for Parameter-Controlled Grading Functions} 
\label{MRDP for controlled GFs}

In some applications the new information may suggest the presence of a hidden control factor whose suggested nature is to be modeled in the least-presuming way as depending on values of certain  "intrinsic" "parameters. For example, modeling the "time-to-failure" in the Reliability Theory using  exponential distribution, its failure rate is often chosen by Maximum Entropy Principle. Using MRDP provides a novel approach to such problems in a more general setup. 

As a representative example, consider a service station where incoming servers stop by the station to pick up groups of customers. With no prior information, IRP suggests the null formula of the service "cost" (say, loading time) of a service batch as proportional to is size: $G(i) = ci$.

The new information suggests that the station's waiting line consists of customers of $K$ different types to be served by only by type-specific incoming servers, so the service batch consists of customers of the same type $k$. It determines the server's capacity $n_k$, and the service cost of a batch of size $i_k \leq n_k$ as an increasing function of the size $F_k(i_k), \quad m_k \leq F_k(i_k) \leq M_k$. 

The observed outflow of served batches and their types suggests modeling the server's  type as a random variable $\kappa$ with stationary probabilities $p_k, k=1, \ldots, K$. The goal of the observer is to find the least-presuming probability distribution of $\kappa$.

By the description, at any server's arrival moment the state of the queue is an element of the chain bundle $W = \{ \vec{i} \} = \{ [i_1, \ldots, i_K] \}$. The expected batch service cost is an additively separable GF on $W$: \quad $F(\vec{i}) = \sum_{k=1}^K p_k F_k (i_k)$.\\

As such, the least-presuming probability distribution of $\kappa$ can be obtained by applying MRDP to $F(\vec{i})$ w.r.t. $G(\vec{i})= c(i_1 + \ldots + i_K)$ on $W$. In \cite{MRDP on Bundles} it led to finding nonnegative $p_1, \ldots, p_K$ maximizing 

\begin{equation}\label{probs}
\sum_{k=1}^K p_k [D_k - (M_k - m_k)\ln{p_k}]
\end{equation},

\noindent under the normalizing constraint $p_1+ \ldots + p_k = 1,$
 
 \noindent where \quad $D_k = \mathcal{D}(F_k \Vert N_k) \vert_{W_k} $.

 Using Lagrange multipliers, it follows that

 $p_k = \exp{[-1 + \frac{D_k - \lambda}{M_k - m_k}]}, \quad k = 1, \ldots, K$,

 \noindent where the $\lambda$ is to be found from the normalizing constraint.

\section{Conclusion}
\label{end}

Relative Divergence (RD) concept and the Maximum Relative Divergence Principle (MRDP) for grading functions (GF) are extended here from totally ordered chains (see \cite{AOME}) to posets of various  structures. MRDP is presented as a mathematical expression of the Insufficient Reason Principle under  prior information (IRP+). The model setup methods are developed and applied. Classic "conditional probability" and "independent events probabilities" formulas are derived from IRP+ expressed as MRDP  posets constructed for that purpose.

Using the "additivity" and "infinum" principles, RD definition formulas were derived for "[$l-g$]" posets connected into serial or parallel circuits. Where the application model's base poset $W$ is an "even-sided split chain", MRDP yielded direct results for the special cases where the involved GFs are "height-dependent". Treating a power set $W = 2^X$ ordered by subset inclusion as an "even-sided split-chain", the definition of RD on that $W$ was developed and used in MRDP analysis. 

In the context of a population $X$ group testing for "positive" subjects, MRDP results enabled the least-presuming (IRP+ based) way to update $F(w)$ - the testing "cost" function of a group $w$ - from the "null" assumption $F(w) = c|w|$ under some extra information. In turn, on the next update stage, the newly observed cost values can be used as constraints for the new MRDP update of $F(w)$.

Another representative application was based on a queue where servers arrive at the station and form their service batches from several waiting lines. "Direct product" of those lines (a "bundle") was set as the base poset $W$. MRDP-suitable models of batch service costs were constructed based on the service contract features as GFs on $W$. MRDP analysis led to direct results. 

In the same context, where the newly observed information suggested service features dependence on a random "type" parameter $\kappa$, the least-presuming distribution of $\kappa$ was obtained.

\section{Acknowledgment}

The author is deeply grateful to prof. Sergei Ovchinnikov (Department of Mathematics of San Francisco State University) for his friendly support, insightful comments and invaluable help in structuring the content and the direction of the research program leading to this paper.


\begin{thebibliography} {12}

\bibitem{Shannon}
C.E. Shannon, A Mathematical Theory of Communication. Bell System Technical Journal, vol. 27, July, October (1948), 623-656.

\bibitem{Dukh}
A. Dukhovny, General Entropy of General Measures, International Journal of Uncertainty, Fuzziness and Knowledge-Based Systems,vol. 10(3) (2002), 213 - 225. 

\bibitem{AOME}
A. Dukhovny,  Axiomatic Origins of Mathematical Entropy: Grading Ordered Sets, arXiv:1903.05240 [math.PR], March 2019

\bibitem{Review}
Jose M. Amigo, Samuel G. Balogh, Sergio Hernandez, A Brief Review of Generalized Entropies, Entropy, 20(2018), 813.  

\bibitem{ElementsIT}, Thomas M. Cover, Joy A. Thomas, Elements of Information Theory, 2nd edition. Wiley, 2006.

\bibitem{K-L}
S. Kullback, R.A. Leibler, On information and sufficiency, Annals of Mathematical Statistics. 22 (1)(1951), 79-86. 

\bibitem{Sinai}
Ya.G. Sinai, On the Notion of Entropy of a Dynamical System. Doklady of Russian Academy of Sciences 124, 768-771 (1959).

\bibitem{Jaynes}
E.T. Jaynes, Information theory and statistical mechanics, Physical Review. 106 (4) (1957), 620 - 630. 

\bibitem{HG}
A. Honda, M. Grabisch, An axiomatization of entropy of capacities on set systems, European Journal of Operational Research, Elsevier, 190 (2) (2008), pp.526-538.

\bibitem{KMR}
I. Kojadinovic, J.-L. Marichal, M. Roubens, An axiomatic approach to the definition of the entropy of a discrete Choquet capacity. Information Sciences 172 (2005), 131 - 153. 

\bibitem{KM}
Kojadinovic, Ivan; Marichal, Jean-Luc, Entropy of bi-capacities,
European Journal of Operational Research, 178(1) (2007), 168-184.

\bibitem{Marichal}
J.-L. Marichal, Entropy of discrete Choquet capacities, Eur. J. of Operations Research, 137 (2002), 612 - 624.

\bibitem{AbDu}
Lev Abolnikov and Alexander Dukhovny, Optimization in HIV Screening Problems.
Journal of Applied Mathematics and Stochastic Analysis, 16:4 (2003), 361-374.

\bibitem{MRDP on P-sets}
A. Dukhovny,  Maximum Relative Divergence Principle for Grading Functions on Power Sets,	arXiv:2207.07099 [math.PR], July 2022

\bibitem{MRDP on Bundles}, Maximum Relative Divergence Principle for Grading Functions on Direct Products of Chains, arXiv:2303.14261 [math.PR], March 2023

\bibitem{P(A|B)}
Alexander Dukhovny, Conditional Probability formula as a consequence of the Insufficient Reason Principle, arXiv: 2507.08040 [math.PR], July 2025

\bibitem{P(AB)}
Alexander Dukhovny, Definition formula for probabilities of independent events 
as a consequence of the Insufficient Reason Principle, arXiv:2508.01147, [math.PR], August 2025  


\end{thebibliography}
\end{document}